\documentclass[11pt]{article}
\usepackage{amsfonts}
\usepackage{float}
\usepackage{jcappub}
\usepackage{graphicx}
\graphicspath{{figures/}}
 \usepackage{caption}
 \usepackage{epstopdf}
 \usepackage{hyperref}
 \usepackage{cleveref}
 \usepackage{pdfpages}
 \usepackage{multicol}
\usepackage{mathrsfs}
\usepackage{amssymb}
\usepackage{textcomp}
\usepackage{amsmath, float}
 \usepackage{times}
 \usepackage{etoolbox}
\usepackage{color}
\usepackage{enumitem}

\title{\boldmath  Probing the cosmic distance duality relation using time delay  lenses }

\author[a,1]{Akshay Rana,\note{Corresponding author.}}
\author[b]{Deepak Jain,}
\author[a]{Shobhit Mahajan,}
\author[a]{Amitabha Mukherjee}
\author[c,d,e]{and R.F.L. Holanda}

\affiliation[a]{Department of Physics and Astrophysics, University of Delhi, Delhi 110007, India}
\affiliation[b]{Deen Dayal Upadhyaya College, University of Delhi,
Sector-3, Dwarka, New Delhi 110078, India}
\affiliation[c]{Departamento de F\'{\i}sica, Universidade Federal de Sergipe, 49100-000, Aracaju - SE, Brazil}
\affiliation[d]{Departamento de Fisica, Universidade Federal de Campina Grande, 58429-900, Campina Grande - PB, Brazil}
\affiliation[e]{Departamento de Fisica, Universidade Federal do Rio Grande do Norte, 59078-970, Natal - RN, Brazil}

\emailAdd{montirana1992@gmail.com}
\emailAdd{djain@ddu.du.ac.in}
\emailAdd{shobhit.mahajan@gmail.com}
\emailAdd{amimukh@gmail.com}
\emailAdd{holanda@uepb.edu.br}

\abstract{The construction of the cosmic distance-duality relation (CDDR) has been widely studied. However,  its consistency with various new observables remains a topic of interest. We present a new way to constrain the CDDR $\eta(z)$ using  different dynamic and geometric properties of strong gravitational lenses (SGL) along with SNe Ia observations. We use a sample of $102$ SGL with the measurement of corresponding velocity dispersion $\sigma_0$ and Einstein radius  $\theta_E$. In addition, we also use a dataset of $12$ two image lensing systems containing the measure of time delay  $\Delta t$ between source images. Jointly these two datasets give us the  angular diameter distance $D_{A_{ol}}$ of the lens. Further, for luminosity distance, we use the $740$ observations from JLA compilation of SNe Ia. To study the combined behavior of these datasets we use a model independent method, Gaussian Process (GP). We also check the efficiency of GP by applying it on simulated datasets, which are generated in a phenomenological way by using realistic cosmological error bars. Finally, we conclude that the combined bounds from the SGL and SNe Ia observation do not  favor any deviation of CDDR and are in concordance with the standard value ($\eta=1$) within $2\sigma$ confidence region, which further strengthens the theoretical acceptance of CDDR. }

\begin{document}
\maketitle
\flushbottom
\section{Introduction}
\label{sec:intro}
The present era of sophisticated observational tools and precision cosmology enables us to  observe  many hitherto unobservable astronomical phenomena. It also provides  a golden opportunity to test the fundamental assumptions which underlie the standard cosmological models. The cosmic distance duality relation (CDDR) or Etherington duality relation  is one such fundamental underlying assumption of observational cosmology. It highlights the link between luminosity distance $D_L$ and angular diameter distance $D_A$:

\begin{equation} \label{CDDRdef}
\eta (z) \equiv \frac{D_L(z)}{D_A(z)(1+z)^2 }=1
\end{equation}

 This relationship between $D_L$ and $D_A$ holds for all metric theories as long as  photon number is conserved in  cosmic evolution and photons follow  null geodesics  \cite{ellis}. Moreover, it acts  as a  fundamental  relation in the background of many cosmological observations such as,  Cosmic Microwave Background (CMB), galaxy clusters \& gravitational lensing. Thus, any departure from it may indicate   the emergence of  new physics or the existence of some unaccounted systematic errors in the observations.
 Several theoretical efforts have been made to study violations of CDDR. For instance, supernova dimming due to the cosmic dust affects the luminosity distance measurement \cite{csaki,lima} and
photon-axion conversion in intergalactic magnetic fields violates photon conservation \cite{avgo,miri}. Similarly, More et. al. studied the modification in the duality relation by introducing the  dilaton and an axion field in a theoretical construction \cite{more} and Piazza \& Schucker (2016) try to understand the behaviour of CDDR in a non-metric theory of gravity \cite{pia}.\par

In recent papers, the CDDR has been tested basically by two approaches. One of them corresponds to less ideal methods that explicitly use some cosmological model in the analysis \cite{basset04,uzan04,avgo09,avgo10,rfl11,rfl16}. The other corresponds to cosmological model independent tests. Several observations have been used to test the CDDR such as $D_A$ obtained from  the gas mass fraction estimate of galaxy cluster, $H(z)$ measurements and baryon acoustic oscillations while  $D_L$ measurements have been taken from SNe Ia  and gamma ray bursts (GRB's) \cite{rfl10,nair11,rfl12,li11,yang13,rflh16}. Moreover, $D_A$ ratio measurements from strong gravitational lenses\cite{rfl16,liao16} and observations of the cosmic microwave background temperature \cite{arana16,ellis13} have also been used for the same purpose.
No significant violation has been observed (see
table 1 in \cite{rfl16}). However, continuing to test the CDDR with different cosmological observations is still important as well as looking for systematic errors  among the previous analysis. \\

In this paper, we present a completely model independent approach to constrain CDDR using strong gravitational lenses (SGL) as standard rulers and SNe Ia as standard candles. In SGL systems when a massive foreground galaxy or cluster (called lens) is positioned in between a source (like QSOs) and the observer, then multiple magnified distorted images of the source  can be seen around the lens. Several dynamical and geometric properties and observables of SGL have been widely used to understand the different aspects of the Universe. For instance,  time delay measurement of lensing images can be used to measure the Hubble parameter $H_0$\cite{ref64,kochanak}. Similarly,  statistical properties of gravitational lenses have also been explored widely to constrain different cosmological parameters such as the cosmological constant \cite{tog,park09,adev03,ffkt}, deceleration parameter \cite{jrgott} and  cosmic curvature \cite{helbig,arana17,xia17}. Here we use the SGL observations as a standard ruler to obtain the angular diameter distance at any redshift. By combining different observations of strong gravitational lensing systems, namely, the Einstein radius \& velocity dispersion measurement of lens galaxy  and the time-delay between the source images, one can obtain a completely model independent $D_A$ . Moreover, this construction of angular diameter distance remain independent of the redshift of the source. Finally, the validity of the cosmic distance duality relation is explored by combining $D_L$ of type Ia supernovae and $D_A$ obtained from the time delay and Einstein radius measurements of lensing systems.

Due to astrophysical complications and instrumental limitations, it is difficult to observe both the luminosity distance $D_L$ and the angular diameter distance $D_A$  of the  same source simultaneously. However, to get an estimate of $\eta$, we need both $D_L$ \& $D_A$ at the same redshift. So in the literature, the use of model independent nonparametric  techniques for this purpose is quite popular. Many such techniques, for instance,  Genetic Algorithm, Nonparametric smoothing technique (NPS),  LOESS \& SIMEX have been used to study the variation of $\eta(z)$ or to check the consistency of different datasets in a non-parametric way \cite{arana16,shafi,nesseris}. In this analysis, we use a model-independent nonparametric technique, Gaussian process, \cite{marina2012,rasmussen2006,zheng,costa} to study the evolution of $\eta$. 

The paper is organized as follows. In the Section 2 we discuss the  method to calculate angular diameter distance using SGL. In Section 3,  we give the details of the datasets used in the analysis. We explain the methodology along with a  model independent non parametric smoothing technique Gaussian Process in section 4. Finally we  summarize the results in section 5.\\

\section{Strong Gravitational Lensing System as a standard ruler}

Strong gravitational lensing (SGL) has emerged as a very powerful observational probe to study the various theories of cosmology and gravity. SGL is the name given to a  phenomenon according to which if a distant object (source) lies behind a mass concentration (lens) (like a galaxy, cluster etc) whose 2D projected matter density is greater then a critical limit $\Sigma_{cr}$  than it can produce multiple images of that distant object. This critical limit $\Sigma_{cr}$  depends upon the mass distribution of the lens and the mutual distances between source, lens and observer. SGL systems observed and detected in SLACS, BELLS, LSD and SL2S surveys have been widely used to put  observational constraints on different cosmological parameters. Working along the similar lines, we present a method by which we can use the Einstein radius and the time delay of the different source images  to obtain the angular diameter distance of lensing mass concentration in the SGL systems.  In the following subsections, we discuss the basic features of Einstein radius, time-delay distance and the methodology to obtain angular diameter distances by combining these quantities.

\subsection{Einstein radius}

Image separation between lenses is one of the directly observable quantities for lensing systems. If the source, lens and observer are aligned along the same line of sight then a ring like structure called, the Einstein ring is formed. For  lenses having a Singular Isothermal Sphere (SIS) mass profile, the image separation between the formed images always remain equal to twice the  Einstein ring's radius ($\theta_E$). It also  depends on the ratio of angular diameter distances (ADD) between lens- source ($D_{A_{ls}}$) and observer-source ($D_{A_{os}}$) \cite{sch06}. Moreover, it is defined as,

\begin{equation}
\theta_E = 4\pi \frac{D_{A_{ls}}}{D_{A_{os}}} \frac{\sigma_{SIS}^2}{c^2}
\label{thetae}
\end{equation}

where c is the speed of light and $\sigma_{SIS}$ is the velocity dispersion of the lens mass distribution. In this way, the quantity of interest is

\begin{equation}
D_{ratio}= \frac{D_{A_{ls}}}{D_{A_{os}}}= \theta_E   \frac{  c^2}{ 4\pi \sigma_{SIS}^2}
\label{dratio}
\end{equation}

\subsection{Time Delay}

The time-delay measurement, $\Delta t$, is another important observational consequence of  SGL\cite{sch06,ref64,sch92,treu10,suyu10}. Unlike the Einstein radius,  it gives a correlation among  three distances: the angular diameter distances between observer \& lens, observer \& source, and lens \& source. It is based on the fact that photons following the null geodesics and originating from a distant source have different optical paths and also pass through dissimilar gravitational potentials. Hence, the time delay is caused by the difference in length of the optical paths and the gravitational time dilation for the ray passing through the effective gravitational potential of the lens.

 Mathematically this quantity is denoted by $\Delta \tau$ and represents the time delay caused due to deflected path with respect to the unperturbed path from source to observer.

\begin{equation}
\label{dtau}
\Delta \tau = \frac{1+z_l}{c} \frac{D_{A_{ol}} D_{A_{os}}}{D_{A_{ls}}} \left[ \frac{1}{2}(\vec{\theta}-\vec{\beta})^2 - \Psi(\vec{\theta}) \right]
\end{equation}

where $ \vec{\theta}$ and $\vec{\beta}$ are the positions of the images and the source respectively, $z_l$ is the lens redshift, and $\Psi$ is the effective gravitational potential of the lens.  Then for a two image lens system (say image A \& B) and  lens having SIS mass profile, the time delay between two source images reduces to \cite{jun}

\begin{equation}
\Delta t=\Delta \tau_A - \Delta \tau_B = \frac{1+z_l}{2c} \frac{D_{A_{ol}} D_{A_{os}}}{D_{A_{ls}}} \left[ \theta_A^2-\theta_B^2 \right]
\label{dt}
\end{equation}

 The quantity $ \frac{D_{A_{ol}} D_{A_{os}}}{D_{A_{ls}}}$ is defined as time-delay distance, $D_{A_{ \Delta t}}$ and given by

\begin{equation}
D_{A_{ \Delta t}}= \frac{D_{A_{ol}} D_{A_{os}}}{D_{A_{ls}}}= \frac{2c\Delta t}{(1+z_l)(\theta_A^2-\theta_B^2)}
\label{dadt}
\end{equation}

\subsection{Angular diameter distance from strong gravitational lensing data}

 Once we obtain the expressions for the different ADD ratios defined as $D_{ratio}$ and $D_{A_{\Delta t}}$, it is  straightforward to calculate angular diameter distance $D_{A_{ol}}$ for lenses. Just by multiplying the estimate of these two distance ratios at the same lens redshift, we get

\begin{equation}
D_{A_{ol}}= D_{A_{\Delta t}} \times  D_{ratio}
\label{daol}
\end{equation}

where $D_{A_{ol}} $ is angular diameter distance of the lens, which is completely independent of source redshift.

\section{Datasets}

The analysis in the paper is based upon the different observational datasets of SGL systems and SNe Ia. In the following sections, we will discuss  the way to find out the angular diameter distance and luminosity distance using  different datasets and their role in the analysis. As discussed in the previous section, in order to calculate the angular diameter distance of a lens, we need the simultaneous measurement of the Einstein radius of the lens and the time delay between lensed images. However, since the discovery of first SGL system Q0957+561 (Walsh.et.al, 1979) thousands of SGL systems have been explored in different surveys [CLASS, PANELS, SDSS, LST etc], but very few could be filtered as a part of a well defined sample. Moreover, it is not possible to find out the time delay distance for all observed lenses,  as its measurement through spectroscopy is dependent on  the variable luminosity of the source. Hence, in our calculation we use the most recent and well defined datasets of Einstein radius and time delay. Further we use a non-parametric method Gaussian process to smoothen these datasets in their redshift range and obtain a model independent measure of angular diameter distance of lens i.e. $D_{A_{ol}}$. The details of the  datasets are as follows;\\

For Einstein radius, we use a set of $102$ lensing galaxies which are part of the Sloan Lens ACS Survey (SLACS), BOSS Emission-Line Lens Survey (BELLS), Lenses Structure and Dynamics (LSD) and Strong Lensing Legacy Survey (SL2S) surveys. This dataset was compiled by Cao et.al (2015) [see Table.1 in ref.\cite{cao15}]. For each lens system, the quantities of interest are the source redshift ($z_s$) \& lens redshift ($z_l$) (determined from SDSS survey), luminosity averaged central velocity dispersion $\sigma_0$ (derived using spectroscopy) and the Einstein radius $\theta_E$ (determined from Hubble Space Telescope (HST) images). The relative uncertainty in   $\theta_E$ is  estimated to be $5$\% across all surveys \cite{cao15}. Moreover, the velocity dispersion for a SIS lens $\sigma_{SIS}$ need not to be the same as the central velocity dispersion $\sigma_0$. Kochanek (1992)\cite{kochanak92} introduced a new parameter $f_E$ such that  $\sigma_{SIS}= f_E \sigma_0$. This parameter $f_E$ compensates for the contribution of dark matter halos in velocity dispersion as well as the systematic errors in measurement of image separation and any possible effect of backgrounnd matter over lenseing systems. All these factors can affect the image separation by up to $20$\%, which can be mimicked by choosing $\sqrt{0.8} <f_E< \sqrt{1.2}$ and   here we choose $f_E$ to be $0.99\pm0.10$ [For details see \cite{ofek03,cao11}]. Finally, using this dataset (see Table.1 in Cao et.al (2015)\cite{cao15}), we obtain $D_{ratio}$, as given in Eq.(\ref{dratio}) and the corresponding error given by \par

   \begin{equation}
\sigma_{D_{ratio}}= D_{ratio}\sqrt{\left(\frac{\sigma_{\theta_E}}{\theta_E}\right)^2 + 4\left(\frac{\sigma_{\sigma_{0}}}{\sigma_0}\right)^2+4\left(\frac{\sigma_{f_E}}{f_E}\right)^2 }
\end{equation}
where $\sigma_{\theta_E}$, $\sigma_{\sigma_0}$ and $\sigma_{f_E}$ are the errors in the measurement of Einstein radius, velocity dispersion and $f_E$ respectively.\\

For the time delay distance we use a dataset of $12$ two-image time delay lensing systems compiled by Balmes \& Corasaniti (2013) \cite{balmes}. In this compilation the quantities of interest are the source redshift $z_s$, the lens redshift $z_l$, the angular positions of both source images with respect to the lens i.e; $\theta_A$ \& $\theta_B$, and the total arrival time difference between both images i.e the  time delay $\Delta t$.  This dataset has been widely used for the estimation of cosmological parameters for different dark energy models \cite{jee,jun,yuen}. We use two-image lensing systems because observationally, such systems are consistent with a simple  SIS mass profile of the lens. Moreover one main reason to consider a SIS mass profile is its simplicity, as it explains the galaxy mass distribution very well\cite{paraficz}  and all parameters like velocity dispersion and time delay can be easily expressed analytically for it. However this selection criterion for two image formation is necessary but not  sufficient to guarantee a SIS mass profile of the lens. So we must include an additional source of error called $\epsilon_{SIS}$ that can aleast account for the several effects which give rise to the observed scatter of individual lenses from a pure SIS mass profile. These effects include the presence of softened isothermal sphere potential, which may decrease the typical image separations,  and the systematic errors in the rms deviation of the velocity dispersion.  According to Cao et al. (2012) \cite{CaoS}, this error can contribute upto $20$\% in the estimation of the time delay distance $D_{A_{\Delta t}}$. Therefore we write $\epsilon_{SIS}= \zeta D_{A_{\Delta t}}$ and choose $\zeta$ to be $0.2$. Now  by applying the error propagation equation on  Eq. (\ref{dadt}) and including possible errors due to the  approximation of  SIS  mass profile of lens, the  net error in the estimation of the time delay  distance $\sigma_{D_{A_{\Delta t}}}$  is given by

\begin{equation}
\sigma_{D_{A_{\Delta t}}}= D_{A_{\Delta t}}\sqrt{\left(\frac{\sigma_{\Delta t}}{\Delta t}\right)^2 + 4\left(\frac{\sigma_{\theta_A} \theta_A}{\theta_B^2 -\theta_A^2}\right)^2+4\left(\frac{\sigma_{\theta_B} \theta_B}{\theta_B^2 -\theta_A^2}\right)^2 + \zeta^2}
\end{equation}

where $\sigma_{\theta_A}$, $\sigma_{\theta_B}$ and $\sigma_{D_{A_{\Delta t}}}$ are the errors in the measurement of positions of source images A \& B and  in measurement of time delay $\Delta t$ respectively.\\

Once  $\sigma_ {D_{ratio}}$ and $\sigma_{D_{A_{\Delta t}}}$ are known, then by using the Eq. (\ref{daol}) we   can easily obtain the error estimate of the angular diameter distance of the lens $D_{A_{ol}}$, given by

\begin{equation}
\sigma_{D_{A_{ol}}}= D_{A_{ol}} \sqrt{\left(\dfrac{\sigma_{D_{A_{\Delta t}}}}{D_{A_{\Delta t}}}\right)^2 + \left(\dfrac{\sigma_{D_{ratio}}}{D_{ratio}}\right)^2}
\end{equation}\\

For the luminosity distance $D_L$, we use  the Joint Light Analysis (JLA) SNe Ia compilation \cite{jla}. It contains a set of $740$ spectroscopically confirmed SNe Ia compiled by the SDSS-II supernova survey \cite{sdss}, SNLS survey \cite{snls} and a few from Hubble Space Telescope (HST) SNe observations \cite{hst} in the redshift region $0.01<z<1.3$. The distance modulus of a SNe Ia is obtained by a linear relation from its light-curve

\begin{equation}
\mu= m_B - (M_B + \alpha \times x_1 - \beta \times c)
\end{equation}

where $m_B$ is the observed rest frame B-band peak magnitude of the SNe, $x_1$ is the time stretching of the light-curve, and $c$ is the supernova color at maximum brightness. These parameters have different value for each SNe Ia and are obtained by fitting the light curves. Thus these are independent of the cosmological model.  $M_B$, $\alpha$, $\beta$ are the nuisance parameters that characterize  the absolute magnitude of the SNe Ia and the shape and the color corrections of the light-curve respectively. These parameters ($M_B$, $\alpha$, $\beta$) are assumed to be constants for all the supernovae. In order to check their dependence on cosmological models, it is desirable to use them as free parameters. However, in recent works it has been observed that $\alpha$ and $\beta$ act like global parameters for a dataset across different cosmological models \cite{subir,jla1,jla2,jla3}. As $\alpha$ and $\beta$ have little effect on our analysis, we will simply adopt the best fit values from Ref. \cite{jla} given as : $\alpha= 1.141\pm0.006$ and $\beta= 3.101\pm0.75$. Once the distance modulus is known then by using the relation $\mu(z)= 5log_{10}(D_L(z)) + 25$, [where $D_L$ is in Mpc] we can easily obtain the luminosity distance $D_L$. The error estimate in $D_L$ is given by

\begin{equation}
\sigma_{D_L}= \frac{\text{ln}(10)}{5}D_L \sigma_{\mu}
\end{equation}\\

\begin{center}
\includegraphics[height=6cm,width=7.5cm]{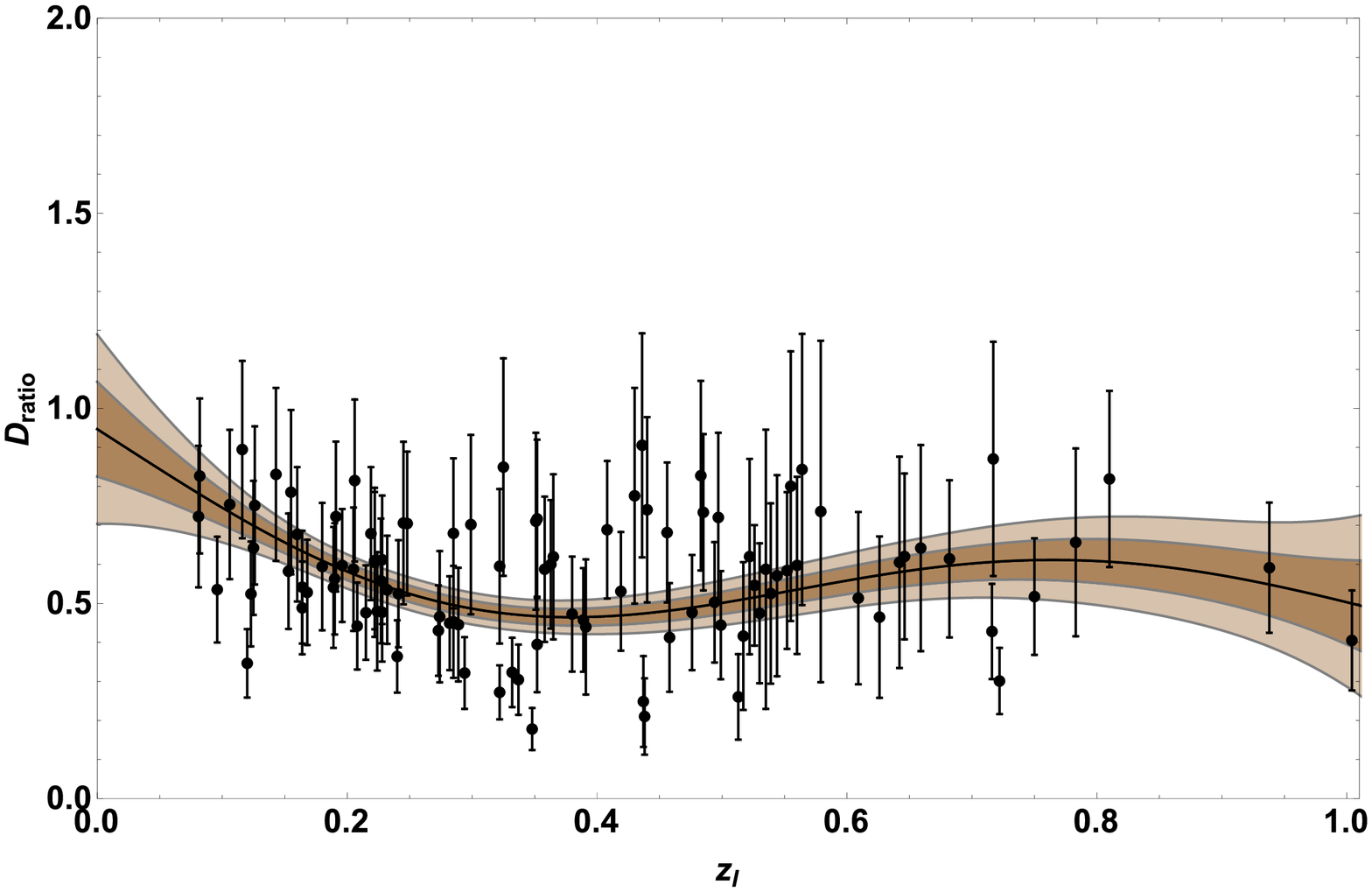}
\includegraphics[height=5.8cm,width=7.5cm]{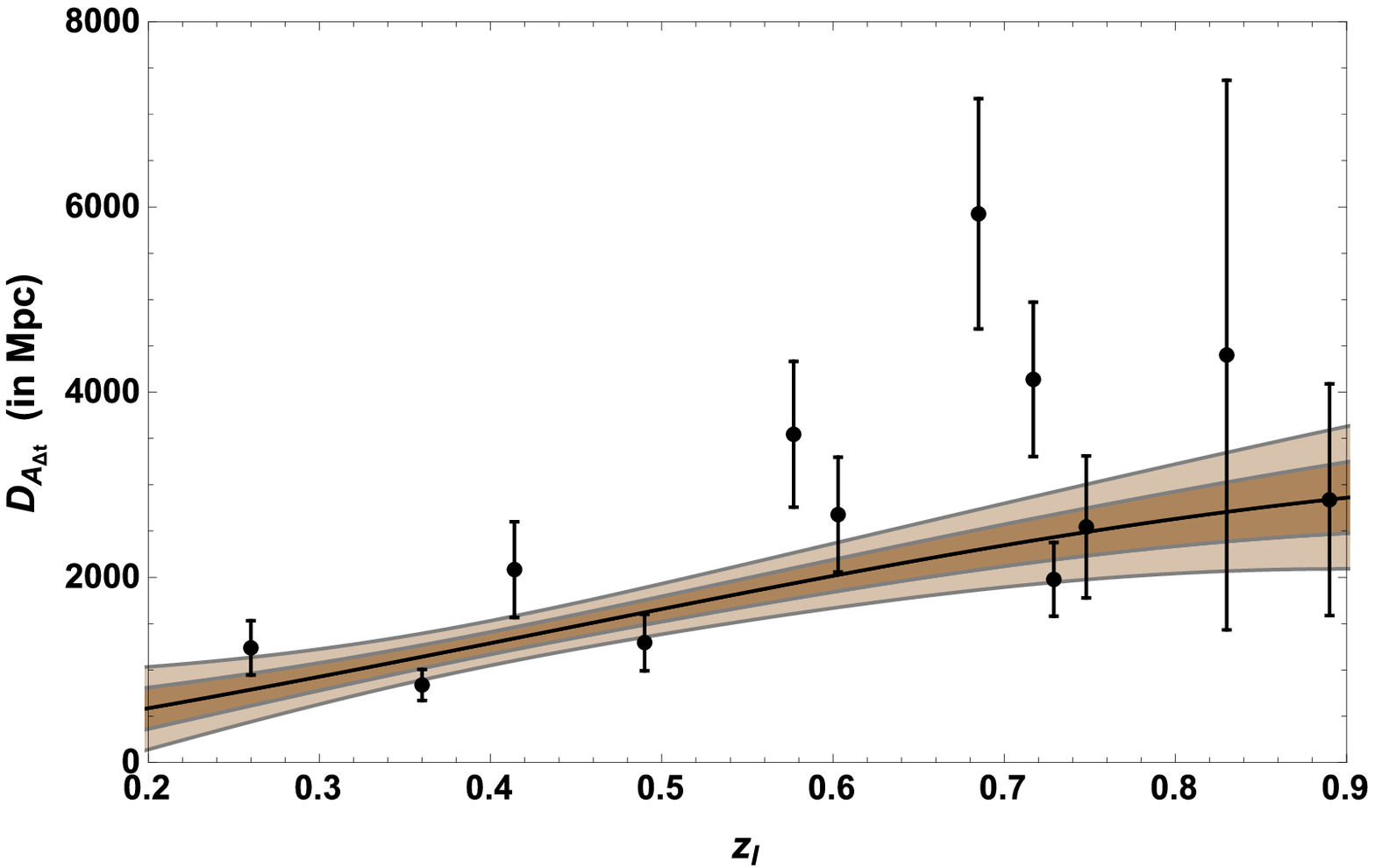}
\captionof{figure}{ \small Black dots with error bars represent the \textbf{[Left plot]} $D_{ratio}=\frac{D_{A_{ls}}}{D_{A_{os}}}$ vs lens redshift $z_l$ data-points for $102$ lensing systems in the redshift region $0.081<z_l<1.004$  and \textbf{[Right plot]} $D_{A_{\Delta t}}=\frac{D_{A_{os}} D_{A_{ol}} }{D_{A_{ls}}}$ vs lens redshift $z_l$ data-points for $12$ time delay distance measurements in the redshift region $0.26<z_l<0.89$. The solid black line represents the best fitted curve to the data obtained using Gaussian Process method along with corresponding $1\sigma$ \&   $2\sigma$ confidence regions in both the plots.}
\label{ER1}
\end{center}

\section{Methodology}

The main objective of this work is to check the validity of the distance duality relation. For this we need the measurement of   luminosity distance and angular diameter distance at the  same redshift, which is not easy  with the available datasets.

 To solve this problem we apply a model independent non-parametric smoothing technique on these datasets and find out the best fit of data with the corresponding error bars. Brief details of this method is given in the next subsection.

\subsection{Gaussian Process}

Gaussian process  is a data interpolation method which can be used to obtain the relationship between a set of input parameters and a set of target, or output parameters. It is a  prominent and widely used non parametric smoothing technique in cosmology. In this method, the complicated parametric  relationship is replaced by  parametrizing a probability model over the data. It is a distribution over functions, characterized by a mean function and a covariance matrix. This method comes with some inherent underlying assumptions like each observation is an outcome of an independent Gaussian distribution belonging to the same population. The outcome of observations at any two redshifts are correlated due to their nearness to each other. This correlation between two points (say $z$ \& $\hat{z}$ ) is incorporated in the technique  through a covariance function given by

\begin{center}
\begin{equation}
k(z, \hat{z})= \sigma_f^2 \exp{-\frac{(|z-\hat{z}|)^2}{2l^2}}
\label{3}
\end{equation}
\end{center}
This square exponential kernel is the simplest and default kernel for GP which comes with two hyperparameters ($l$ \& $\sigma_f$). The length-scale $l$   determines the length of the 'wiggles' in the smoothing function, while output variance $\sigma_f$ fixes the average distance of function away from its mean. The values of these hyperparameters are calculated  by maximizing the corresponding marginal log-likelihood probability function of the distribution. We can also use this information to reconstruct the outcome or observable value at different redshifts using a multivariate Gaussian probability distribution.  We use this method to reconstruct the value of angular diameter distance $D_{A_{ol}}$ and luminosity distance $D_L$. For more details of GP refer to \cite{marina2012, rasmussen2006}

\begin{center}
\includegraphics[height=6cm,width=7cm,scale=4]{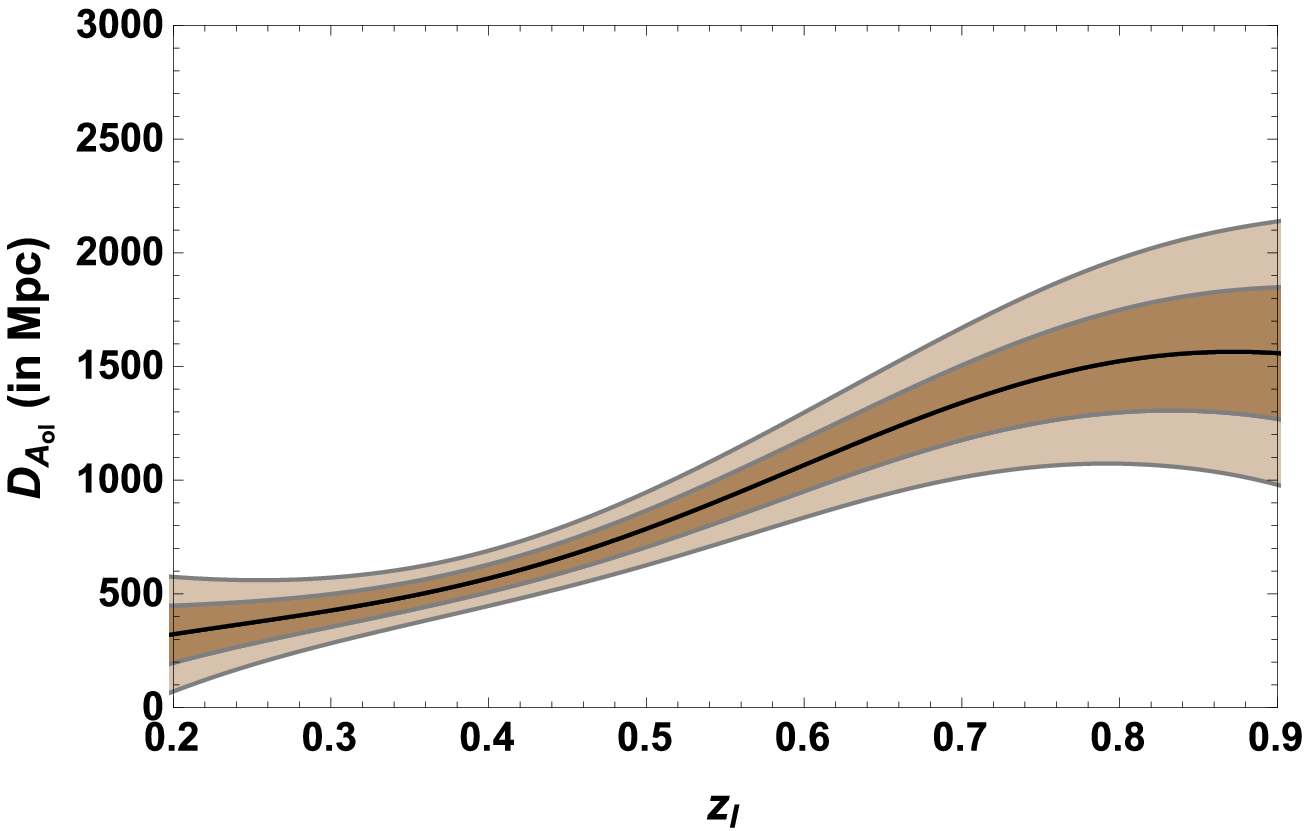}
\includegraphics[height=6cm,width=7cm,scale=4]{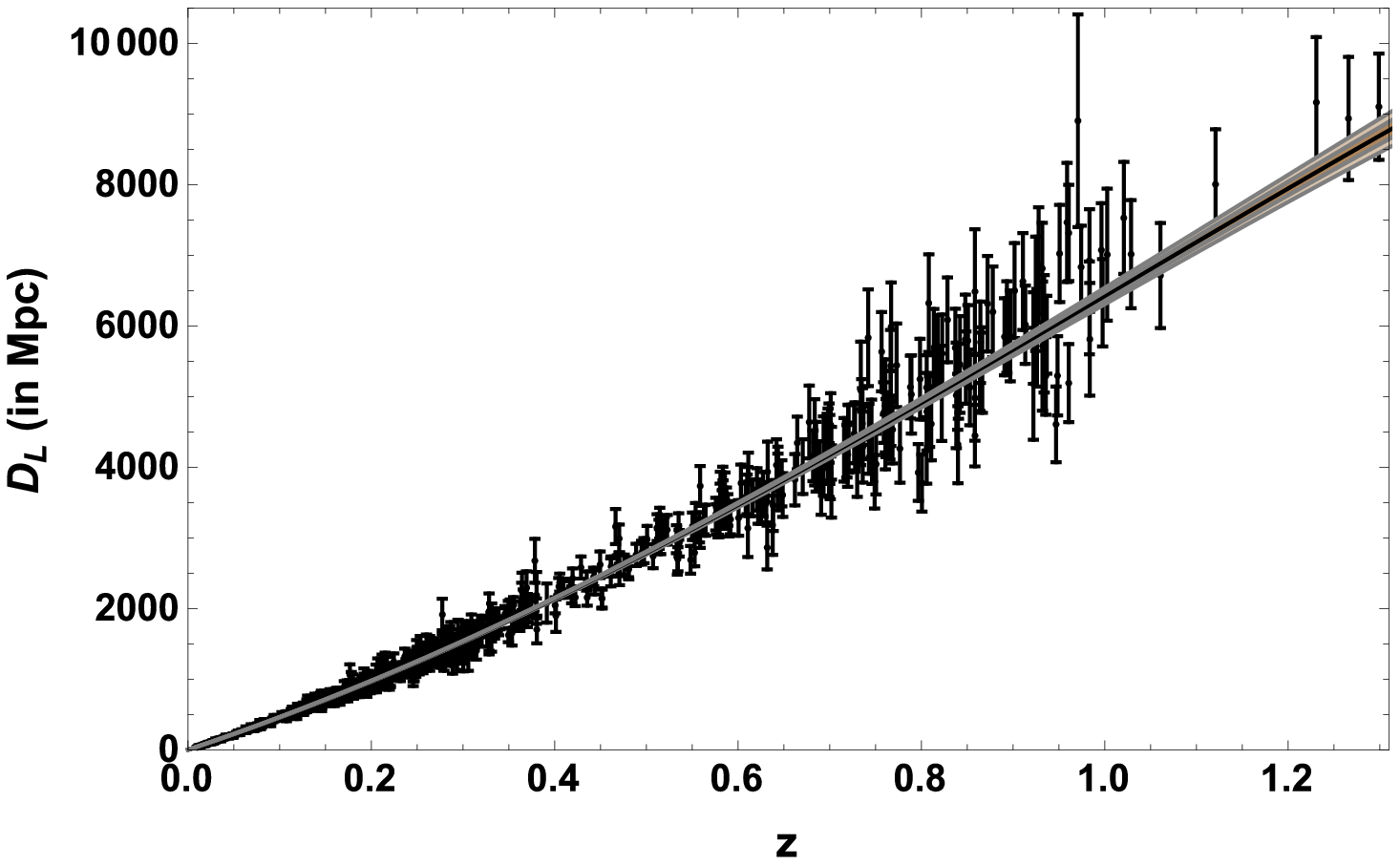}
\captionof{figure}{\small  \textbf{[Left plot]}: This figure gives an estimate of angular diameter distance of lenses $D_{A_{ol}}$ in the redshift range $0.2<z_l<0.9$ obtained by combining smoothed plots of Fig \ref{ER1}. \textbf{[Right plot]} Black dots with error bars represent the  luminosity distance $D_L$ vs  $z$ data-points of JLA compilation having $740$ SNe Ia in the redshift region $0.01<z<1.3$. The solid black line represents the best fitted curve to the data obtained using Gaussian Process method along with corresponding $1\sigma$ \&   $2\sigma$ confidence regions. }
\label{ERT}
 \end{center}

\subsection{Generation of mock datasets}

 Gaussian process is one of the most established smoothing technique in cosmology. In order to check the efficiency of GP, that is  whether it recovers  the underlying behavior of $\eta$ or not, we generate two simulated datasets based on  known and realistic cosmologies. The first dataset is based on the underlying assumption of the  $\Lambda CDM$ model, which implies that  $\eta$ should always remain equal to $1$. In the other simulated dataset, we assume a violation of CDDR by using an underlying fiducial model,   $\eta(z)= 1+ \frac{z}{(1+z)^2}$. We generate $100$ datapoints for each assumed model. Further, to introduce a realistic error  on these datapoints we use a phenomenological approach. Here we utilize the redshift dependence of the error of the real dataset. For this purpose, we use  the real dataset of $\eta$ formed by using the luminosity distance and angular diameter distance measure as shown in Fig.\ref{ERT}. The details of the method used are explained in Ref. \cite{congma,arana16}.
 
Once the mock datasets are generated, the GP technique is used to check whether it is capable enough to reconstruct the underlying behaviour of the dataset. On analysing Fig.\ref{MOCK}, It is observed that for both the models (i) where CDDR holds i.e, $\eta=1$  and (ii) in which $\eta = 1$ is violated  i.e, $\eta(z)= 1+ \frac{z}{(1+z)^2}$, GP  recovers the assumed fiducial behaviour  within a $1\sigma$ confidence region. This analysis gives us confidence that GP is a very efficient model independent non parametric smoothing technique.

\begin{center}
\includegraphics[height=6cm,width=7.5cm,scale=4]{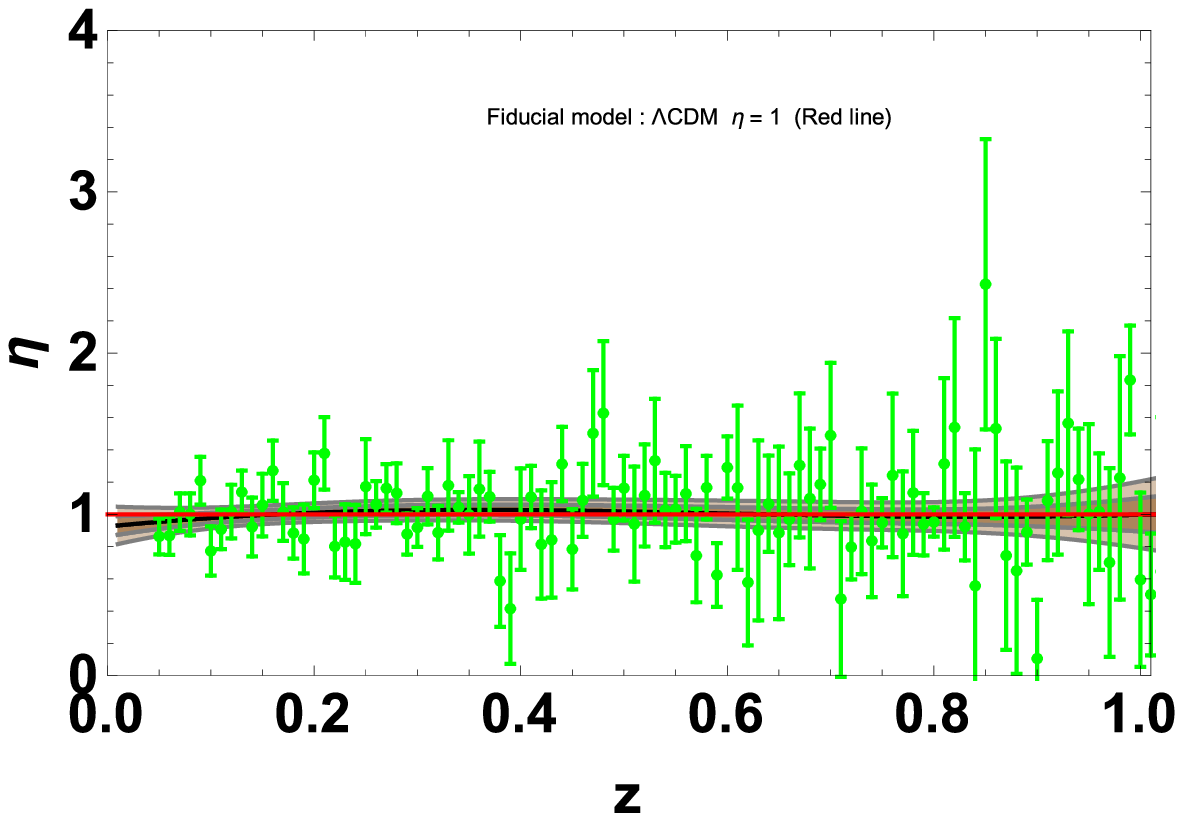}
\includegraphics[height=6cm,width=7.5cm,scale=4]{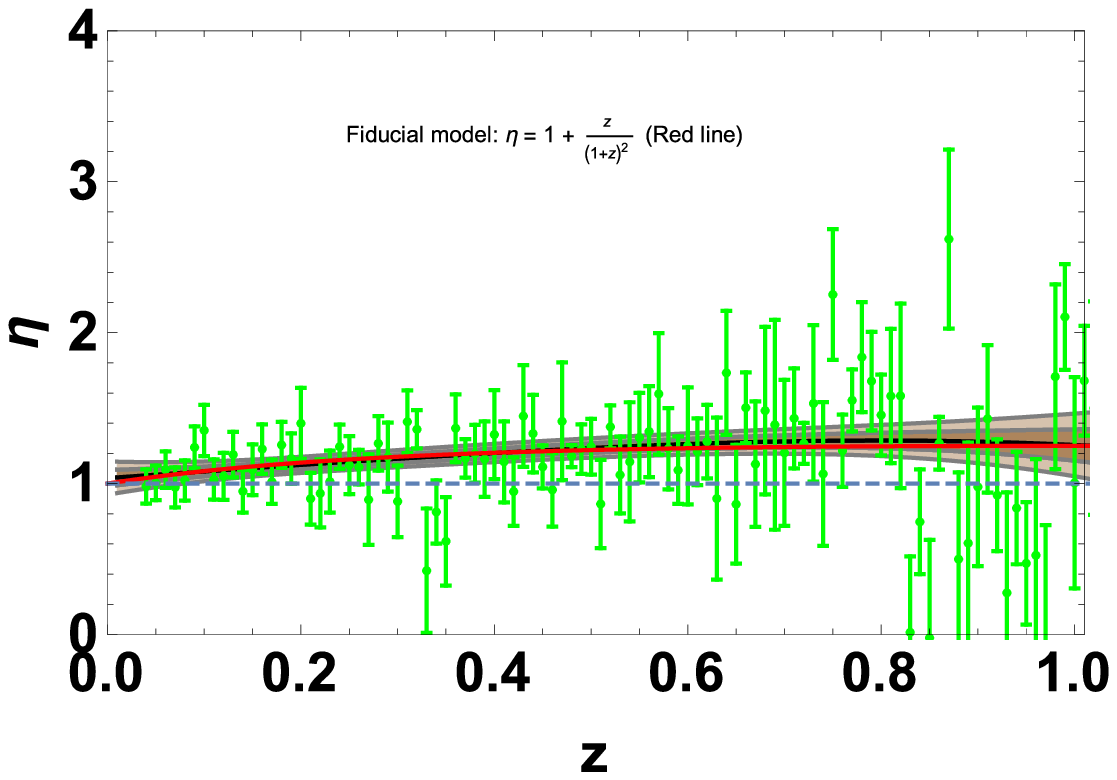}
\captionof{figure}{\small  \textbf{[Left plot]}:  Green points represents the 100 mock datapoints of $\eta$ vs $z$ simulated by using the fiducial model $\Lambda CDM$ i.e. $\eta=1$ (Red line). \textbf{[Right plot]}:  Green points represents the 100 mock datapoints of $\eta$ vs $z$ simulated by using the fiducial model $\eta(z)=1+\frac{z}{(1+z)^2}$ (Red line). While black line represents the reconstructed plot along with $1\sigma$ \&   $2\sigma$ confidence regions after applying GP for both the plots. Moreover, dashed line represents the $\eta=1$.
}
\label{MOCK}
 \end{center}

\subsection{Reconstruction of the CDDR }

Cosmic distance duality relation (CDDR) is an interconnection between the cosmological luminosity distance and angular diameter distance, as defined in equation \ref{CDDRdef}. For CDDR to hold, $\eta$ should be equal to $1$. To test it, we need an estimate of $D_A$ and $D_L$ at the same redshift. We apply GP smoothing on the $D_{ratio}$ dataset obtained by using Einstein radius \& velocity dispersion measurements of strong gravitational lensing system [see left panel in Figure \ref{ER1}] and on the $D_{A_{\Delta t}}$ dataset formed by using the time delay measured between the two-image strong gravitational lenses [see right panel in Figure \ref{ER1}]. Interestingly, from Eq (\ref{daol}) we have seen that both these measure of $D_{ratio}$ and $D_{A_{\Delta t}}$ jointly give us an estimate of the angular diameter distance $D_{A_{ol}}$ of lenses in SGL systems, which remains independent of the source redshift. Hence just by multiplying the reconstructed curves of $D_{ratio}$ and $D_{A_{\Delta t}}$, obtained by applying GP, we get a model independent measure of the angular diameter distance $D_{A_{ol}}$. [see left panel in Figure \ref{ERT}]. However to avoid the extrapolation of the smoothing function, we limit the measure of   $D_{A_{ol}}$ to the redshift range $0.2<z<0.9$. Furthermore, on applying the GP on  the JLA SNe Ia dataset, we also obtain the measure of luminosity distance in the required  redshift range.  [See right panel Figure \ref{ERT}].  Finally, this continuous estimate of luminosity distance $D_L$ and angular diameter distance $D_{A_{ol}}$ gives us the smoothed curve of the CDDR parameter $\eta$ along with the $1\sigma$ \& $2\sigma$ confidence bands [see Figure \ref{ETA}]. This curve includes the standard $\eta= 1$ line within a $2\sigma$ confidence region and indicates a consistency of CDDR within the  redshift region $0.2<z<0.9$. The confidence regions are obtained by using the error propagation method for  the luminosity distance $\sigma_{D_L}$ and the angular diameter distance $\sigma_{D_{A_{ol}}}$, given by

\begin{equation}
\sigma_{\eta}= \eta \sqrt{\left(\dfrac{\sigma_{D_L}}{D_L}\right)^2 + \left(\dfrac{\sigma_{D_{A_{ol}}}}{D_{A_{ol}}}\right)^2}
\end{equation}\\

\begin{center}
\includegraphics[height=6cm,width=9cm,scale=4]{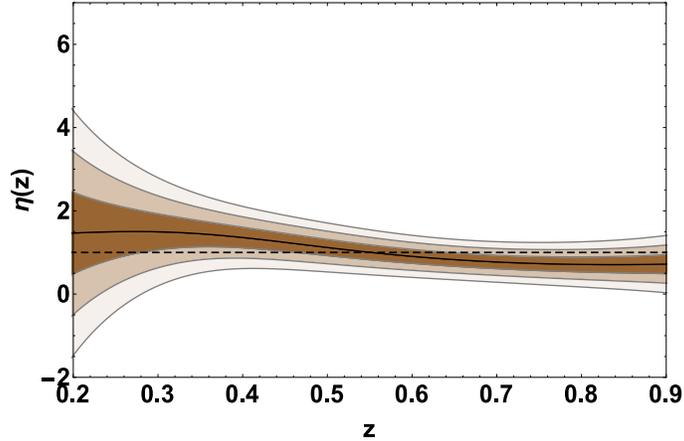}
\captionof{figure}{\small Solid black line represents the reconstructed smooth curve of CDDR parameter $\eta$ vs $z$ in the redshift range $0.2<z<0.9$. The standard $\eta= 1$ line is represented by the dashed black line. The inward to outward bands represent the $1\sigma$,  $2\sigma$  \&  $3\sigma$ confidence regions.}
 \label{ETA}
\end{center}

\section{Discussion}

The cosmic distance duality relation is one of the most fundamental relations in  cosmology. It is interesting  to check the validity of this relation by using the luminosity distance ($D_L$) and the angular diameter distances ($D_A$). However to check the consistency of $\eta$ we need both $D_L$ and $D_A$ at the same redshift. With the present limitations of the observational capabilities, it is not possible to obtain  both distances simultaneously. Hence in the present scenario it seems a good alternative to use the distance estimates coming from different sources. In  recent times SNe Ia has emerged as the  best estimate of luminosity distance so we use  JLA  compilation  of $740$ SNe Ia for the measure of $D_L$. Moreover, the measurement of angular diameter distance at higher redshifts always remains a challenging task due to observational difficulties to identify the standard rulers. Even in the present era of precision cosmology the best estimates of $D_A$ come from the Galaxy clusters and Baryonic Acoustic Oscillation (BAO). However  since the discovery of the first strong gravitational lenses, it has emerged as a very powerful cosmological probe to study  cosmology. Different geometric and dynamical analysis of these lenses provide us an estimate of  the different angular diameter distance relations, which have been used to constrain  different cosmological models and parameters.

  Many  efforts have been made to test the CDDR in recent past using SGL. Liao et. al. (2015) \cite{liao}, Holanda et. al (2016) \cite{rfl16} and Fu et.al (2017)\cite{fu} use the image separation and velocity dispersion  dataset of SGL to calculate the Angular Diameter Distance (ADD) ratio  and combine it with different estimates of  the luminosity distance from SNe Ia (Union2.1 or JLA catalouge) and GRBs. However in all these analysises the ADD ratio is used, which   depends on the redshift of the source in SGL systems. We don't have a measure of luminosity distance at  sufficiently high redshifts compared to the  source redshift of SGLs.  Hence a lot of SGL observations get  rejected at the very first step of analysis. However, Liao et. al. used the GRB's for $D_L$ \cite{liao}, which can act as a strong observable to probe the  Universe at higher redshifts but it's credibility  as a standard ruler is still a question of debate.  In several works related to the study of different cosmological models, it has already  been observed that the collective results from the Einstein radius and the time delay  are more sensitive to the cosmological parameters compared to their separate analysis \cite{jun,yuen,paraficz}.  Till now the collective effect of these two different features of SGL have not been used to constrain CDDR. This motivates  us to propose a completely new and model independent way to estimate the ADD using these dynamical and geometrical features of SGL, which is completely independent of the source redshift and further enables us to use most of the SGL systems for analysing CDDR .

 In this work we use a new method to estimate the angular diameter distance of SGL systems. In this methodology, the Einstein radius is obtained from the deflection angle and the time delay between the different images collectively determine from the angular diameter distance of lenses in SGL systems. We use the Einstein radius measure of $102$ SGL systems to obtain $D_{ratio}$ and $12$ time delay measures of two image lensed systems to calculate $D_{A_{\Delta t}}$. Further, by applying Gaussian process, we  find the variation of these distances over the redshift range  $0.2<z<0.9$. By multiplying these two distances, we obtain the angular diameter distance between observer and lens, $D_{A_{ol}}$. Further, the luminosity distances $D_L$ are obtained from  JLA SNe Ia  dataset . Using these two datasets ($D_{A_{ol}}$ and $D_L$), we constrain the CDDR parameter ($\eta$).\\

 The efficiency of the statistics that is being used in such type of analysis is always a matter of concern. Hence to ensure that GP is competitive enough to identify the real underlying behaviour of the dataset, we simulate two mock datasets having realistic errors. In one simulated dataset, the  fiducial model supports the CDDR $\eta=1$, while in the other dataset  violates this relation $\eta(z)= 1+ \frac{z}{(1+z)^2}$. After applying GP on both the  datasets [see fig. \ref{MOCK}], we observe that reconstructed plots completely recover back the underlying models in simulated datasets within a  $1\sigma$ confidence level. This analysis further strengthens our confidence in  the efficiency of GP and improves the final results. Our results are as follows:\\

$\bullet$ From Fig. \ref{ETA}, it is clear that the reconstructed CDDR parameter ($\eta$) derived by using luminosity distances from SNe Ia JLA compilation and angular diameter distances from SGL is in concordance with the standard $\eta=1$ line within  the  $2\sigma$ confidence region. It also shows its consistency with other results obtained by using different distance indicators like BAO, CMB, Galaxy cluster, radio galaxy and GRBs [for comparison see Table.1 in \cite{rfl16}].

$\bullet$  Most of the lensing galaxies are early type massive elliptical galaxies and in several studies it has been observed that a SIS mass profile is a well tested and strongly supported mass profile for such lenses \cite{fuku92,kochanak96,helbig99,legg10,ruff11}. However, Schwab et.al.(2010) tested the sensitivity of cosmological parameters over lens profile and stellar velocity dispersion \cite{sch}. They observed that even a small departure in these measures could bring significant  changes in the constraints on cosmological parameters. Moreover, if we also include the effect of anisotropy in  velocity dispersion, slope of mass profile and luminosity in this analysis, then for best fit mean values of these parameters the distance ratio $D_{ratio}= \frac{D_{A_{ls}}}{D_{A_{os}}}$ could further decrease upto 12\% \cite{xia17,rasanan,caos}. This may increase the mean value of $\eta$  approximately by the same amount.

$\bullet$ Though, in this analysis we used all  prescribed  errors that can compensate for the assumption of lens mass distribution (SIS  profile) and velocity dispersion  as suggested by Cao et.al (2015) \cite{cao15}, we still believe that independent knowledge of mass profile of lensed galaxies and their dynamical  structure may bring a significant change in lensing studies. The spatially resolved dynamical analysis of nearby galaxies from the surveys like SAURON \cite{ens,capp} and high resolution ground based spectroscopy with large telescope may reduce these systematic and statistical errors and provide independent priors for the gravitational lenses.

$\bullet$ The  ongoing gravitational lensing dedicated surveys like the COSMOGRAIL project \cite{cosmograil} and others like SLACS, BELLS, LSD \& SL2S  along with upcoming new and powerful breed of sky surveys like DES, LSST, JDEM ,VST ATLAS, Pan-STARRS, EUCLID mission, TMT \& E-ELT are expected  to find  thousands of new SGLs. New observations with much better precision will act as a game-changer and help to establish SGL  as one of the most effective probes to constrain CDDR as well as to understand the universe. However, with present constraints on our resolving capabilities of celestial objects, it seems difficult to estimate the time delay and velocity dispersion for the same SGL system simultaneously. This is because the systems for which time delay is measurable are surrounded by bright quasars and for such lenses, getting the correct estimate of velocity dispersion is difficult due to higher contamination of background source luminosity.  In such cases this work provides a  statistically precise and  model independent way to obtain the ADD for gravitational lens systems.

\section*{Acknowledgments}
Authors are thankful to an anonymous referee for useful comments which substantially improved the paper. A.R. acknowledges support under a CSIR-SRF scheme (Govt. of India, F.No. 09/045(1345/2014-EMR-I)) and also thanks IUCAA Resource Centre, University of Delhi, for providing research facilities. R.F.L.H. is supported by INCT-A and CNPq (No. 478524/2013-7;303734/2014-0).

\end{document}